\documentclass{article}

\usepackage[nonatbib,final]{nips_2018}

\usepackage[utf8]{inputenc} 
\usepackage[T1]{fontenc}    
\usepackage{hyperref}       
\usepackage{url}            
\usepackage{booktabs}       
\usepackage{amsfonts}       
\usepackage{nicefrac}       
\usepackage{microtype}      

\usepackage{graphicx}

\usepackage{algorithm}
\usepackage{algpseudocode}

\usepackage[all]{xy}

\usepackage{amsbsy}
\usepackage{amsmath}

\usepackage{bbm}

\newcommand{\bfX}{\boldsymbol{X}}
\newcommand{\bfy}{\boldsymbol{y}}
\newcommand{\bfc}{\boldsymbol{c}}

\newcommand{\beginsupplement}{%
        \setcounter{table}{0}
        \renewcommand{\thetable}{S\arabic{table}}%
        \setcounter{figure}{0}
        \renewcommand{\thefigure}{S\arabic{figure}}%
     }

\title{Using permutations to assess confounding in machine learning applications for digital health}

\author{
  Elias Chaibub Neto$^\ast$ \\
 \And
  Abhishek Pratap \\
  \And
  Thanneer M Perumal \\
  \And
  Meghasyam Tummalacherla  \hspace{0.5cm}  Brian M Bot  \hspace{0.5cm} Lara Mangravite  \hspace{0.5cm} Larsson Omberg \\
  Sage Bionetworks, Seattle, WA 98121 \\
  \texttt{$^\ast$elias.chaibub.neto@sagebase.org} \\
}

\begin{document}

\maketitle

\begin{abstract}
Clinical machine learning applications are often plagued with confounders that can impact the generalizability and predictive performance of the learners. Confounding is especially problematic in remote digital health studies where the participants self-select to enter the study, thereby making it challenging to balance the demographic characteristics of participants. One effective approach to combat confounding is to match samples with respect to the confounding variables in order to balance the data. This procedure, however, leads to smaller datasets and hence impact the inferences drawn from the learners. Alternatively, confounding adjustment methods that make more efficient use of the data (e.g., inverse probability weighting) usually rely on modeling assumptions, and it is unclear how robust these methods are to violations of these assumptions. Here, rather than proposing a new approach to control for confounding, we develop novel permutation based statistical methods to detect and quantify the influence of observed confounders, and estimate the unconfounded performance of the learner.  Our tools can be used to evaluate the effectiveness of existing confounding adjustment methods. We illustrate their application using real-life data from a Parkinson's disease mobile health study collected in an uncontrolled environment.
\end{abstract}

\section{Introduction}

Machine learning algorithms have been increasingly used as diagnostic tools in biomedical research\cite{gulshan2016,esteva2017,golden2017}. The widespread availability of smartphones and other health tracking devices generates high volumes of sensor data, and makes machine learning uniquely well posed to impact clinical research using digital health tools. In clinical applications, gender, age, and other demographic characteristics of the study participants often play the role of confounders. Confounding is particularly prevalent in mobile health studies run under uncontrolled conditions outside clinical and laboratory settings, where we have little control over the demographic and clinical characteristics of the cohort of participants that self-select to participate in a study.

In the context of predictive modeling, we define a confounder as a variable that causes spurious associations between the features and response variable. In machine learning applications, the presence of confounding can lead to ambiguous inference and poor generalizability of models. Confounding is usually present when the joint probability distribution of the confounder and response variables is different in the data available to develop the learner (which we from now on denote as the “development dataset”) relative to the population where the learner will be applied (denoted as the “target population”)\cite{rao2017}. For example, consider a diagnostic application where most cases are old aged while most controls are young, but where age is not associated with disease status in the target population (e.g., the target population is composed of older patients only). If the classifier can more efficiently detect age-related signals than disease-related signals, then it will likely perform poorly when deployed in the target population.

Confounding adjustment is an active area of research in machine learning. The goal is to prevent an algorithm from learning the confounding signal. Since any variable that confounds the feature-response relationship has to be associated with both the features and the response, most of the methods proposed in the literature can be divided into two approaches: (i) methods  that remove the association between the confounder and the response; or (ii) methods that remove the association between the confounder and the features. A standard example of the first approach is to match subjects from the development data in order to obtain a subsample that more closely resembles the target population. This strategy, however, results in a smaller number of participants to train and evaluate the machine learning algorithm, and, in highly unbalanced situations, might lead to the exclusion of most of the participants from the analyses. Alternative methods that make more efficient use of the data include inverse probability weighting approaches\cite{linn2016,rao2017}, which weight the training samples in order to make model training better tailored to the target population. A canonical example of the second approach is to separately regress each feature on the confounders, and use the residuals as the predictors in the machine learning algorithm. Other approaches that do not fall into categories (i) or (ii) include penalized learners\cite{li2011} and backdoor adjustment\cite{landeiro2016}.

In this paper, we present statistical methods to detect and quantify the influence of observed confounders,  and to estimate the actual (i.e., unconfounded) predictive performance of a learner. We use a large Parkinsons digital health study cohort to illustrate how our methods can be used to evaluate the effectiveness of standard confounding adjustment methods.

\section{Methods}

We adopt restricted permutations\cite{good2000,rao2017} to isolate the contribution of the confounder from the predictive performance of a learner. The key idea is to shuffle the response data within the levels of a categorical/ordinal confounder (as illustrated in the Supplementary Figure \ref{fig:stratifiedperm}) in order to destroy the direct association between the response and the features while still preserving the indirect association due to the confounder. Algorithm \ref{alg:stratifiedShuffling} describes the procedure for an arbitrary performance metric, $m$ (such as the area under the receiver operating characteristic curve, AUC, or root mean square error).

\begin{algorithm}
\caption{Restricted Monte Carlo permutation null distribution for performance metric $m$}\label{alg:stratifiedShuffling}
\begin{algorithmic}[1]
\State \textbf{Input}: Number of permutations, $b$. Development data set feature matrix, response vector, and confounder vector, $\bfX$, $\bfy$, $\bfc$. Training and test set indexes, $i_{train}$, $i_{test}$
\State Split $\bfX$, $\bfy$ and $\bfc$ into training and test sets
\For{$i = 1, 2, \ldots, b$}
  \State $\bfy^{\ast}_{train} \leftarrow \mbox{RestrictedShuffle}(\bfy_{train}, \bfc_{train})$, and $\bfy^{\ast}_{test} \leftarrow \mbox{RestrictedShuffle}(\bfy_{test}, \bfc_{test})$
  \State Train a machine learning algorithm on the $\bfX_{train}$ and $\bfy_{train}^{\ast}$ data
  \State Evaluate the algorithm on the $\bfX_{test}$ and $\bfy_{test}^{\ast}$ data
  \State Record the value of the performance metric, $m^\ast_i$, on the shuffled data
\EndFor
\State \textbf{Output}: $m^\ast_1$, $m^\ast_2$, \ldots, $m^\ast_b$
\end{algorithmic}
\end{algorithm}

Building upon the restricted permutation null distribution, we developed two statistical tools to deal with confounding. First, we estimate the ``unconfounded” predictive performance of a learner by building a mapping from the restricted permutation null to the standard permutation null (where the standard permutation null distribution is generated by shuffling the labels in the usual unconstrained manner). As fully described in the Supplement, for any performance metric that can be expressed as a (generalized) U-statistic\cite{hoeffding1948,lehmann1951,serfling1980,delong1988} (e.g., AUC), or expressed as a simple average (e.g., mean square error, mean absolute error, and classification accuracy), we have that an asymptotic estimate of the unconfounded performance metric is given by,
\begin{equation}
\hat{m}_u \, = \, (m_o - a_{\hat{\pi}^{\ast}}) (s_{\hat{\pi}^{\ast\ast}}/s_{\hat{\pi}^{\ast}}) + a_{\hat{\pi}^{\ast\ast}}~,
\label{eq:correctedmetric}
\end{equation}
where $m_o$ represents the uncorrected metric value; $a_{\hat{\pi}^{\ast}}$ and $s^2_{\hat{\pi}^{\ast}}$ represent the sample average and variance of the restricted permutation null; and $a_{\hat{\pi}^{\ast\ast}}$ and $s^2_{\hat{\pi}^{\ast\ast}}$ represent the analogous quantities for the standard permutation null. It is important to point out that we don’t view the unconfounded metric estimation as an adjustment method (in the sense that it does not prevent an algorithm from learning the confounding signal in the first place). It simply quantifies the amount of response signal learned by the algorithm, after the algorithm has had a chance to learn both confounding and response signals.

Second, by noticing that the location of the restricted permutation null provides a natural measure of the amount of confounding signal learned by the algorithm, we adopt the average of the restricted permutation null as a test statistic, and develop a statistical test to compare the hypotheses,
\begin{align}
H_0^c &: \mbox{the algorithm has not learned the confounding signal~,} \\
H_1^c &: \mbox{the algorithm has learned the confounding signal~,} \nonumber
\end{align}
and detect confounding learning per se. In the Supplement we show that, under $H_0^c$, the average of the restricted permutation null distribution is asymptotically distributed as a $N(a_{\hat{\pi}^{\ast\ast}}, s^2_{\hat{\pi}^{\ast\ast}}/n)$ distribution, where $n$ represents the test set sample size.

For the AUC metric, additional analytical results are available. It is well known\cite{bamber1975,mason2002} that the standard permutation null can be approximated by,
\begin{equation}
N(0.5 \, , \, (n_n + n_p + 1)/(12 \, n_n \, n_p))~,
\label{eq:standnullapprox}
\end{equation}
where $n_n$ and $n_p$ represent the number of negative and positive labels in the test set. Hence, $a_{\hat{\pi}^{\ast\ast}} \approx 0.5$ and $s^2_{\hat{\pi}^{\ast\ast}} \approx (n_n + n_p + 1)/(12 \, n_n \, n_p)$ and the estimator in (\ref{eq:correctedmetric}) becomes,
\begin{equation}
auc_u = (auc_o -  a_{\hat{\pi}^{\ast}})(n_n + n_p + 1)/(12 \, n_n \, n_p \, s_{\hat{\pi}^\ast}) + 0.5~,
\label{eq:auccorrectionformula}
\end{equation}
the null distribution under $H_0^c$ reduces to $N(0.5, (n_n + n_p + 1)/(12 \, n_n \, n_p \, n))$, and,
\begin{equation}
p = 1 - \Phi\big((a_{\hat{\pi}^{\ast}} - 0.5)/[(n_n + n_p + 1)/(12 \, n_n \, n_p \, n)]^{1/2}\big)~,
\label{eq:confoundingpval}
\end{equation}
corresponds to the confounding p-value, where $\Phi(.)$ represents the c.d.f. of standard normal variable.

\section{Real data illustrations}

A key practical application of our tools is to evaluate if an adjustment method is working as expected. This is important in practice since most of these methods rely on assumptions, and it is generally unclear how robust they are to violations of these assumptions. Here we illustrate the application of our tools to two confounding adjustment methods: sample matching, and approximate inverse probability weighting (IPW) based on the propensity score\cite{propensityscore1983}.

\begin{figure}[!h]
  \centering
  \centerline{\includegraphics[width=3.8in, bb = 0 10 420 300]{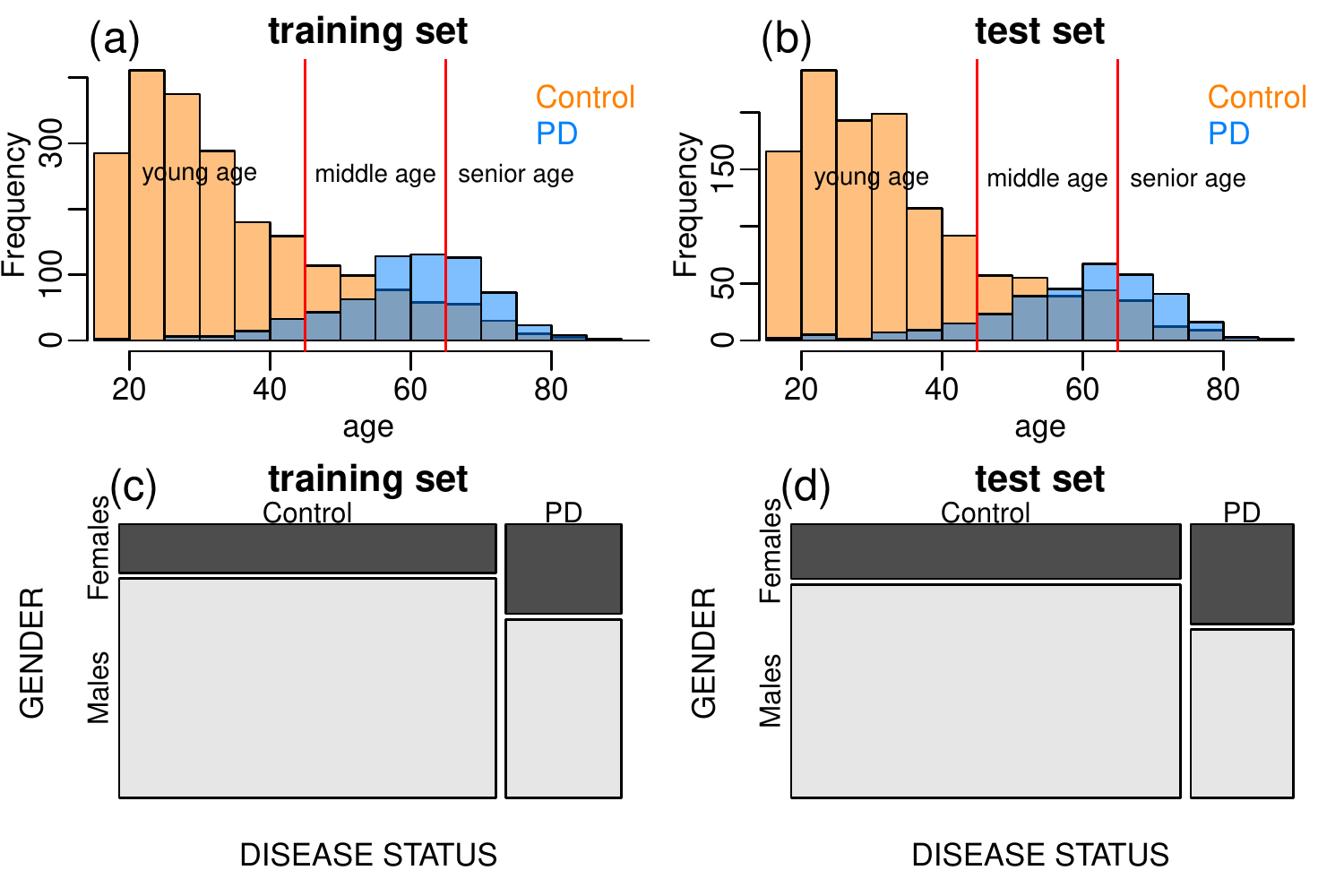}}
  \caption{Age and gender associations in the mPower data. Panels a and b show that, for both training and test sets, the age distributions of PD and control participants have reduced overlap, with control participants being usually younger than PD participants leading to a strong association between age and disease status. Panels c and d present mosaic plot of disease status by gender, showing again an association between these variables (with a larger proportion of female participants in the PD group than in the control group). The training and test sets are composed, respectively of 658 and 331 cases and 2144 and 1255 controls.}
  \label{sfig:agegenderassociations}
\end{figure}

Our development data was collected in a digital health study on Parkinsons disease\cite{bot2016,trister2016} and consist of features generated from 30 second inertial sensor readings captured during walking. We focused on walking, as walking patterns are influenced by age and gender\cite{ko2011} in addition to Parkinson’s disease. The development data was split into training and test sets with similar joint distributions for the age, gender, and disease status (Supplementary Figure \ref{sfig:agegenderassociations}). We apply the adjustment methods to both training and test sets, and the analyses are based on a combined gender/discretized age\footnote{While, in theory, we can only perform restricted permutations using categorical/ordinal confounders, in practice we can discretize and evaluate continuous confounders as well. Clearly, if the discretization is too coarse the discretized confounder might not be able to fully capture the association between the confounder and the response, and we might end up underestimating the amount of confounding learned by the algorithm. In practice, one should experiment with distinct discretizations, as illustrated in Supplementary Figure \ref{sfig:agediscretization}.} confounder with levels: young male, young female, middle age male, middle age female, senior male, and senior female.

\begin{figure}[h]
  \centering
  \centerline{\includegraphics[width=\linewidth, bb = 0 20 720 210]{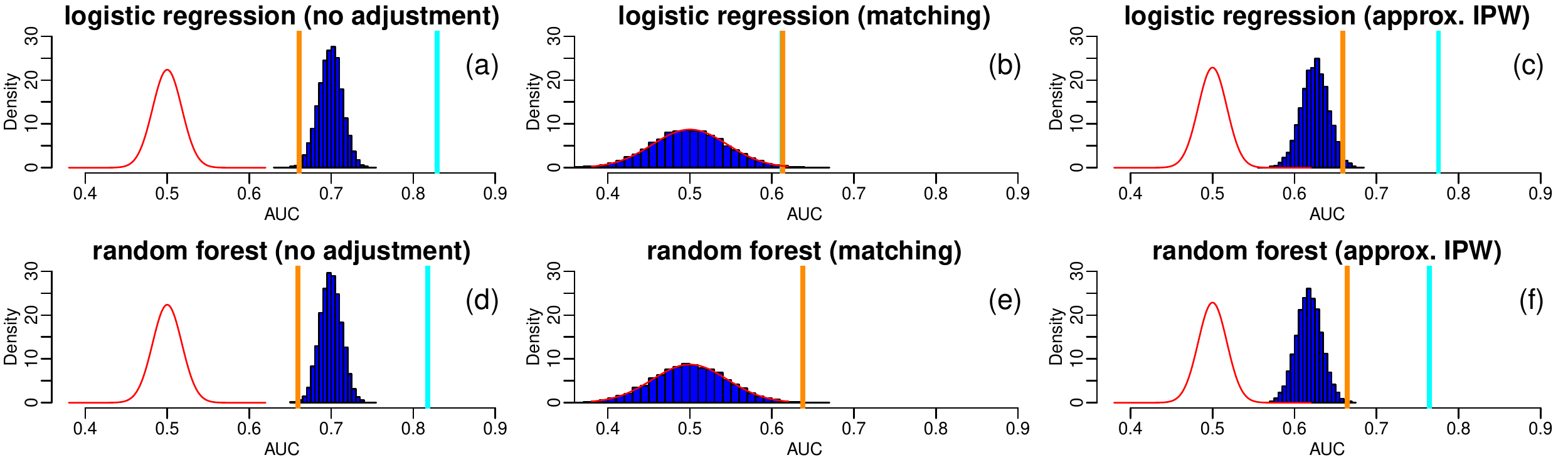}}
  \caption{Comparison of confounding adjustment methods.}
  \label{fig:adjustmentcomparison}
\end{figure}

Figure \ref{fig:adjustmentcomparison} shows the results based on logistic regression (top panels) and random forest (bottom panels) classifiers. In all panels, the blue histograms represent the restricted permutation null distributions generated by Algorithm 1, the red curves represent the normal approximation for the standard permutation null distribution presented in equation 3, the orange line shows the unconfounded estimate of AUC computed using equation \ref{eq:auccorrectionformula}, and  the cyan line represents the observed AUC.

For the sake of comparison, panels a and d report the results when no confounding adjustment is performed. Both logistic regression and random forest classifiers are clearly learning confounding signal since the restricted permutation nulls are centered around 0.7, and the confounding test p-values (equation \ref{eq:confoundingpval}) are highly significant ($p<10^{-16}$). Hence, the high AUC scores (cyan lines above 0.81) reflect the classifiers' ability to detect both disease and confounding signals, while the unconfounded estimates (orange scores around 0.66) are considerably more modest.

Panels b and e show the results based on a matched subset of participants. The fact that the restricted permutation nulls are centered around 0.5, and closely match the standard permutation null density (red curve), suggests that matching effectively prevented the classifier from learning the confounding signal ($p<0.51$ and $p<0.58$, respectively) and that the classifiers are only learning the disease signal. As expected, the observed and unconfounded AUC scores match each other closely in this situation. Finally, note that the much larger spread of the null distributions (in comparison to panels a and d) is due to the smaller test set available after matching.

\begin{figure}[!h]
  \centering
  \centerline{\includegraphics[width=4.2in, bb = 0 0 650 520]{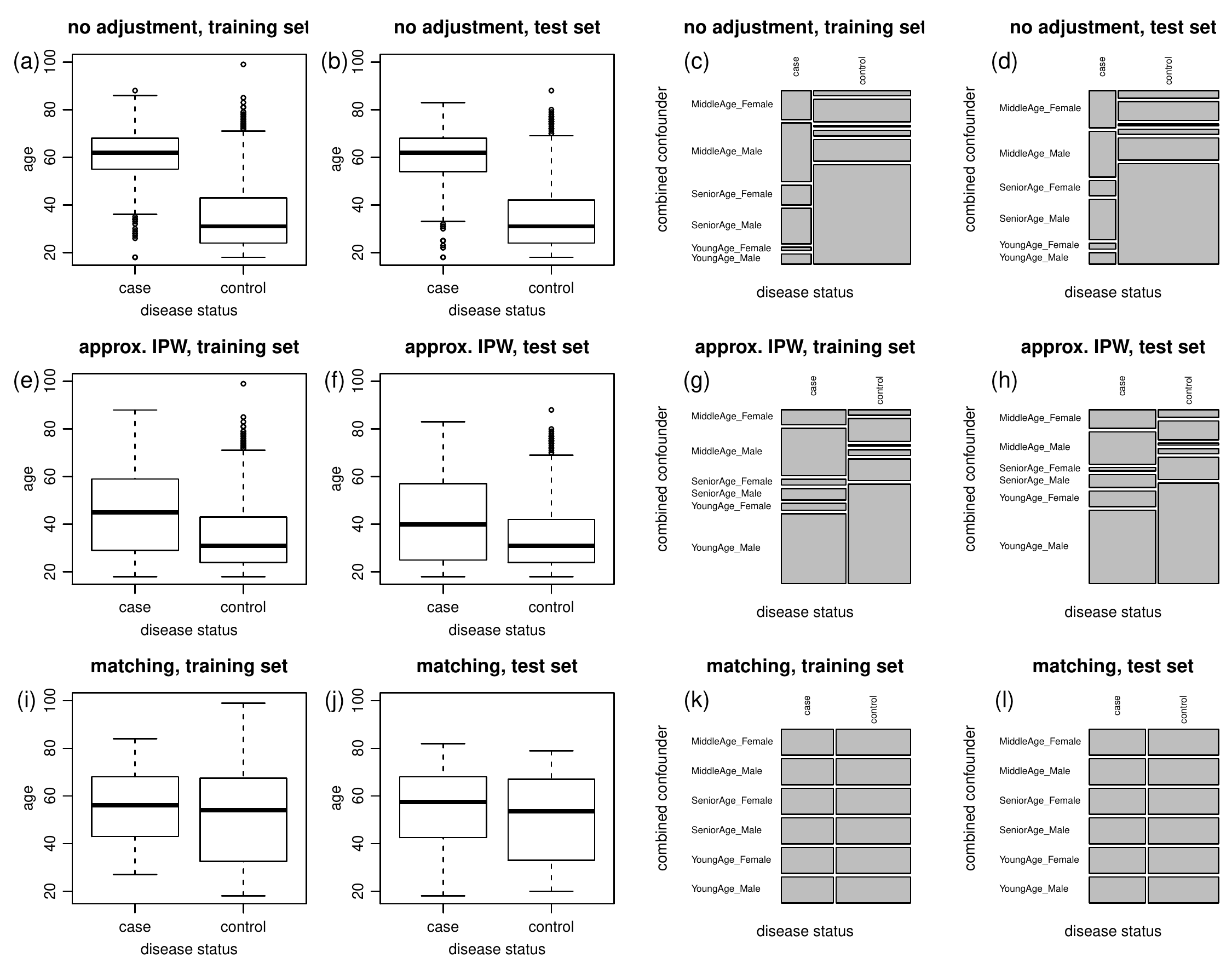}}
  \caption{Checking confounder balancing achieved by the approximate IPW and the matching approaches. For the sake of comparison, panels a to d show the results for the original data. Panels a and b show boxplots of age by disease status for the training and test sets. Panels c and d show the respective mosaic plots for the combined age/gender confounder versus the disease status (age was discretized into young, middle, and senior age categories). Panels e to h show the respective plots for the augmented training and test sets generated by the approximate IPW method. While the method clearly improved the balance (in comparison to the results in the top panels), it still did not manage to generate truly well balanced training and test sets. Panels i to l show the results for the matching approach. Now, the mosaic plots (panels k and l) show a perfect balance for the combined age/gender confounder versus the disease status. The boxplots also show better (although not perfectly) balanced age distributions. (Note that the balance is not perfect because we performed the matching using the discretized age, instead of the original age variable.)}
  \label{sfig:failuretobalance}
\end{figure}

Panels c and f report the results for the approximate IPW approach. This method makes use of the entire data set and attempts to prevent confounding learning by weighting the samples according to the inverse of their estimated propensity scores (i.e., the conditional probability that a participant has the disease given its gender and age). While several approaches have been proposed in the literature for the estimation of propensity scores\cite{lee2010,pirracchio2014}, here, we adopt the most commonly used method based on logistic regression. The panels show that the approximate IPW approach managed to reduce the amount of confounding (the blue histograms are closer to 0.5 compared to panels a and d). However, it didn’t remove it completely ($p < 10^{-16}$). This suggests that the estimated inverse probability weights did not generate a well balanced augmented data set. (Supplementary Figure \ref{sfig:failuretobalance} confirms this is indeed the case.) Most likely, the reason for this suboptimal performance is that propensity score estimation using logistic regression makes the strong assumption that the observed associations between the confounders and disease labels can be well described by the logistic function. This example illustrates how the violation of a parametric modeling assumption can lead to an inefficient confounding adjustment.

\section{Final remarks}

Digital health enabled diagnostic systems have the potential to provide low cost remote diagnostic tools to underserved communities that lack easy access to medical care. However this opportunity cannot be fully realized without (i) efficient approaches to combat confounding (without which we run the risk of making spurious inferences from the data) and (ii) rigorous methods to evaluate these adjustment methods. The tools proposed in this paper address the second need.

To the best of our knowledge, the use of restricted permutations in the context of predictive modeling has only been leveraged by\cite{rao2017}. These authors, however, use restricted permutations to test if a machine learning algorithm has learned the response signal in the presence (or absence) of confounders, but not to detect and quantify confounding learning per se, as proposed in this paper.

For the sake of clarity, our illustrations focused on the case where confounder and response are associated in the development set but not in the target population. We can, however, still apply our methodology when response and confounder are known to be associated in the target population but have a different joint probability distribution compared to the development data. Section 9 of the Supplement provides an illustrative example based on synthetic data.

We also conducted a simulation study to evaluate the statistical properties of the confounding test (Section 10 of the Supplement). Our simulations show reasonable statistical power for the range of parameters investigated in our experiments, and well-controlled type I error rates.

Finally, we point out that while this paper has focused on digital health applications, the proposed tools can be more generally applied to any other areas impacted by confounders.

\beginsupplement

\clearpage

\section{Supplementary Figures}

\begin{figure}[!h]
  \begin{center}
  \centerline{\includegraphics[width=5in, bb = 0 50 1000 500]{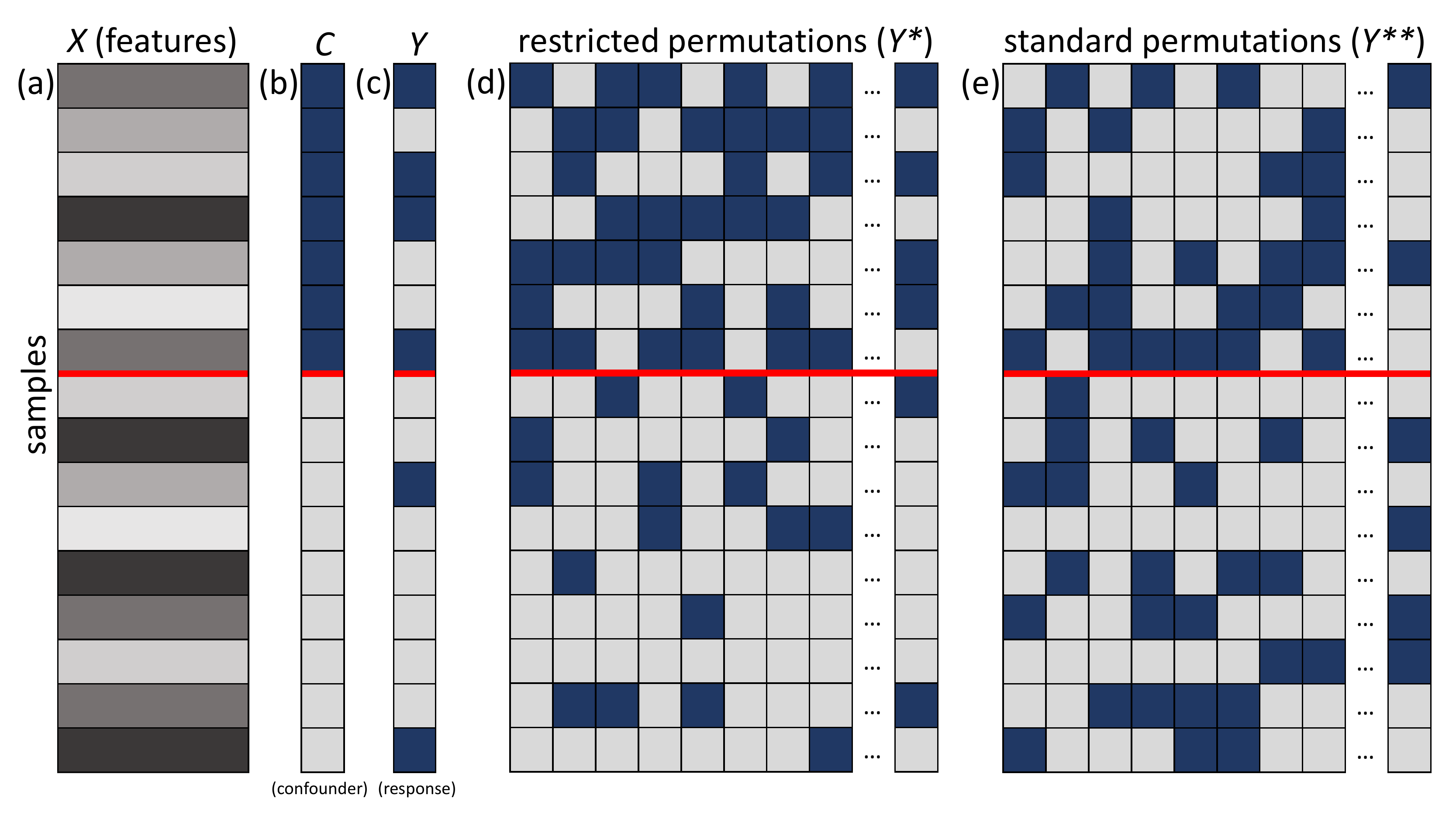}}
  \end{center}
  \caption{Panels a, b, and c, represent the features ($X$), confounder ($C$), and response ($Y$) data, respectively. In this cartoon example, we have 16 samples, and both $C$ and $Y$ are binary (light and dark cells represent 0 and 1 values, respectively). The confounder vector (panel b) was sorted, and the red line splits the data relative to the levels of $C$ (i.e., the top 7 samples have confounding value 1, while the bottom 9 have confounding value 0). Note that in panel c we have 4 positive response values (dark cells) above the red line, and 2 below it. Panel d illustrates the restricted permutation scheme. Each column shows a distinct permutation. In all permutations, we still have 4 dark cells above the red line and 2 below it. The restricted permutations destroy the association between $Y$ and $X$, while still preserving the association between $Y$ and $C$. Panel e illustrates the standard permutation scheme, where we shuffle the response values freely across the entire response vector (now, each column is no longer constrained to have 4 dark cells above the red line and 2 below it). The standard permutations destroy the association between $Y$ and $C$ and between $Y$ and $X$.}
  \label{fig:stratifiedperm}
\end{figure}

\begin{figure}[!h]
  \centering
  \centerline{\includegraphics[width=4.2in]{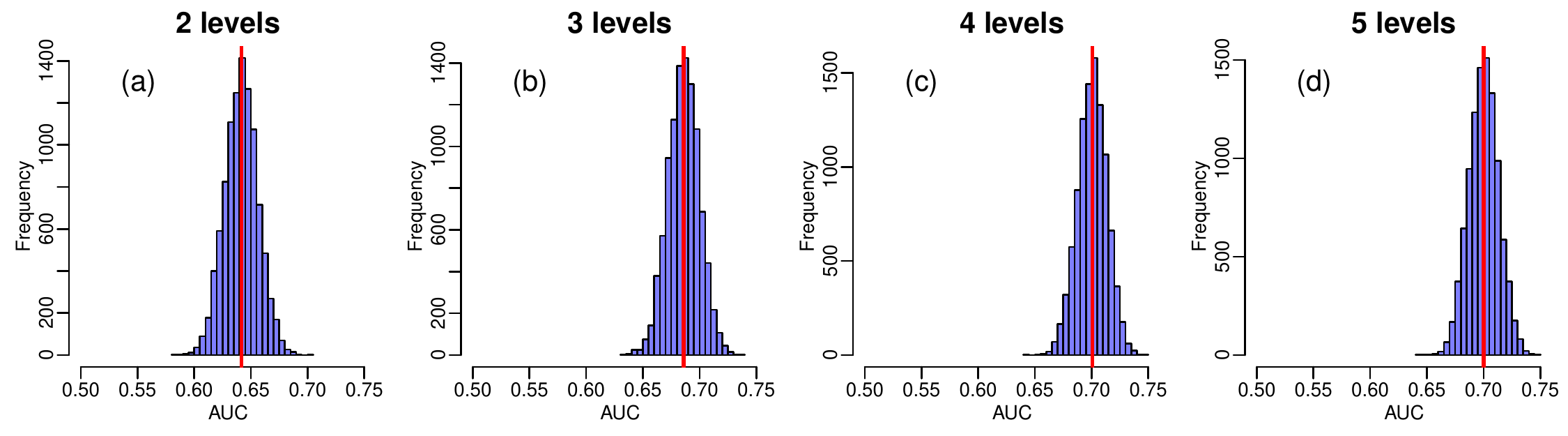}}
  \caption{Here, we illustrate how the granularity of the discretization can influence the amount of confounding signal detected by the restricted permutation null distribution. As pointed out in the main text, if the discretization is too coarse the discretized confounder might not be able to fully capture the association between the confounder and the response, and we might underestimate the amount of confounding learned by the algorithm. Here, we experiment with distinct discretizations of the age confounder, namely, categorizing age into 2, 3, 4, and 5 levels. (These level categorizations are given, respectively, by the following age ranges $\{[18, 58], [59, 99]\}$, $\{[18, 44], [45, 65], [66, 99]\}$, $\{[18, 35], [36, 50], [51, 65], [66, 99]\}$, and $\{[18, 30], [31, 45], [46, 60], [61, 75], [76, 99]\}$.) Inspection of the results suggest that the discretization based on 4 levels seems to be enough for this feature set, as increasing the discretization to 5 levels does not shift the restricted null. Clearly, splitting age into 2 levels is not enough since the restricted permutation null is located at much lower AUC values, showing that a fair amount of confounding signal was not captured by this coarse discretization. Splitting age into 3 levels still seem to miss some of the association, as we obtain a slightly stronger confounding signal using 4 levels. (In the paper illustrations, however, we continue to use the 3 level categorization, as it already captures enough age signal.)}
  \label{sfig:agediscretization}
\end{figure}

\clearpage

\section{The unconfounded metric estimate}

The observed metric $m_o$ captures the contributions of both response and confounder learning. In order to estimate the ``unconfounded" value, $m_u$, we need to determine what value would the observed performance metric have assumed, had the response variable not been associated with the confounder. In other words we need to map a value sampled from a distribution where the response and confounder are associated to a distribution where they are not. To this end, we construct a mapping from the restricted permutation null distribution (where the association between the response and the confounder is preserved) to the standard permutation null (where this association is removed).

\begin{figure}[!h]
  \centerline{\includegraphics[width=2.0in, bb = 30 0 400 120]{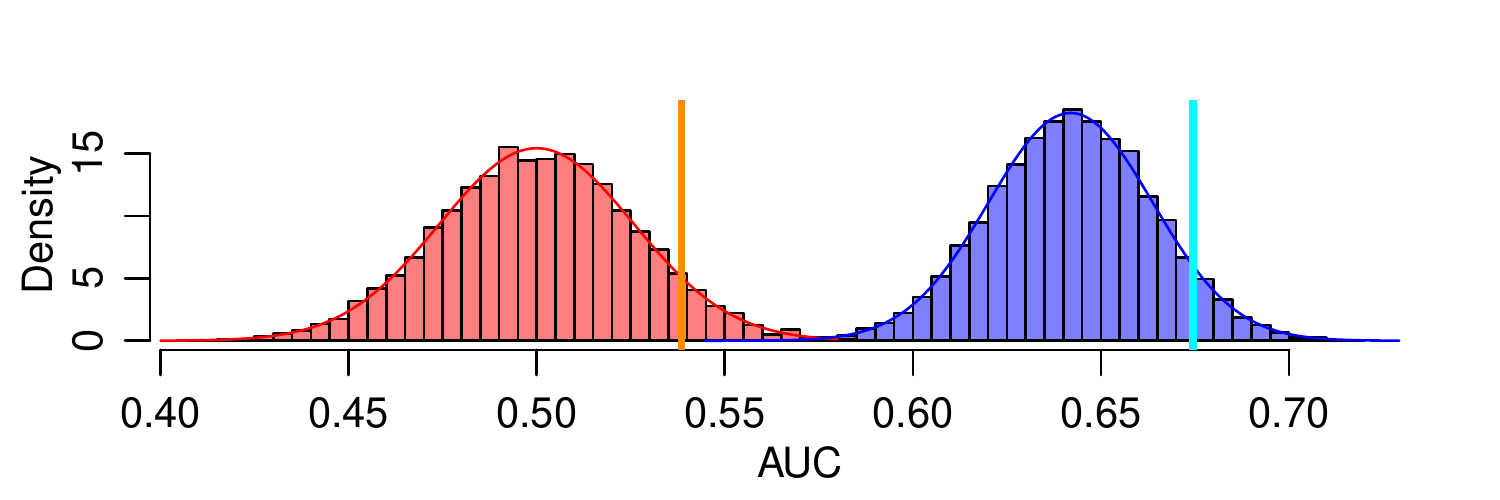}}
  \caption{The figure shows an example of the restricted (blue) and standard (red) permutation null distributions for the AUC metric. The cyan line represents the observed AUC value ($m_o$), while the orange line shows the unconfounded estimate ($\hat{m}_u$). Note that the tail probabilities to the right of the cyan and orange lines are the same (i.e., the p-values are preserved).}
  \label{fig:matchedpvals}
\end{figure}

Let $F_{\pi^\ast}$ and $F_{\pi^{\ast\ast}}$ represent, respectively, the restricted and standard permutation null distributions, and $\hat{F}_{\hat{\pi}^\ast}$ and $\hat{F}_{\hat{\pi}^{\ast\ast}}$ represent the respective Monte Carlo versions of these permutation distributions. An obvious mapping would be to define $m_u = m_o - a_{\hat{\pi}^{\ast}} + a_{\hat{\pi}^{\ast\ast}}$, where $a_{\hat{\pi}^{\ast}}$ and $a_{\hat{\pi}^{\ast\ast}}$ correspond, respectively, to the sample mean of $\hat{F}_{\hat{\pi}^\ast}$ and $\hat{F}_{\hat{\pi}^{\ast\ast}}$. This mapping, however, only focus on the means and fails to take into consideration the different spreads of the restricted and standard permutation null distributions. Ideally, we should define a mapping that accounts for the entire probability distributions. Therefore, we define and estimate the unconfounded metric $m_u$ by equating $F_{\pi^{\ast\ast}}(m_u)$ to $F_{\pi^\ast}(m_o)$,
\begin{equation}
F_{\pi^{\ast\ast}}(\hat{m}_u) \, = \, F_{\pi^\ast}(m_o) \; \Leftrightarrow \; \hat{m}_u \, = \, F_{\pi^{\ast\ast}}^{-1}(F_{\pi^\ast}(m_o))~.
\label{eq:correctedmetricperm}
\end{equation}
Note that, equating $F_{\pi^\ast}(m_o)$ to $F_{\pi^{\ast\ast}}(m_u)$ is equivalent to equating the p-values, as illustrated in Figure \ref{fig:matchedpvals}.

In general, $\hat{F}_{\hat{\pi}^\ast}$ and $\hat{F}_{\hat{\pi}^{\ast\ast}}$ are unknown distributions. However, because popular performance metrics such as the mean square error, mean absolute error, and the classification accuracy correspond to averages, while metrics such as the AUC correspond to generalized U-statistics (DeLong, DeLong and Clarke-Pearson 1988; Lehmann 1951), we have that the distribution of these statistics can be well approximated by Gaussian distributions when the test set is large enough (due to central limit theorems associated with averages, and to the asymptotic normality of (generalized) U-statistics (Hoeffding 1948, Serfling 1980)). Hence, in practice, we will often be able to approximate $\hat{F}_{\hat{\pi}^\ast}$ and $\hat{F}_{\hat{\pi}^{\ast\ast}}$ by,
\begin{equation}
\hat{F}_{\hat{\pi}^{\ast}} \, \approx \, N(a_{\hat{\pi}^{\ast}} \, , \, s^2_{\hat{\pi}^{\ast}})~, \hspace{0.5cm} \hat{F}_{\hat{\pi}^{\ast\ast}} \, \approx \, N(a_{\hat{\pi}^{\ast\ast}} \, , \, s^2_{\hat{\pi}^{\ast\ast}})~,
\label{eq:approxgaussian}
\end{equation}
where $s^2_{\hat{\pi}^{\ast}}$ and $s^2_{\hat{\pi}^{\ast\ast}}$ correspond, respectively, to the sample variances of $\hat{F}_{\hat{\pi}^\ast}$ and $\hat{F}_{\hat{\pi}^{\ast\ast}}$, and $a_{\hat{\pi}^{\ast}}$ and $a_{\hat{\pi}^{\ast\ast}}$ represent, as before, the respective sample means. (The blue and red densities on top of the histograms in Figure \ref{fig:matchedpvals} correspond, respectively, to the normal approximations in (\ref{eq:approxgaussian}).) Now, by replacing $F_{\pi^\ast}$ and $F_{\pi^{\ast\ast}}$ in equation (\ref{eq:correctedmetricperm}) by the approximate Gaussian distributions in (\ref{eq:approxgaussian}) we have that,
\begin{align}
\hat{F}_{\hat{\pi}^{\ast\ast}}(\hat{m}_u) \, &\approx \, \Phi\big((\hat{m}_u - a_{\hat{\pi}^{\ast\ast}})/s_{\hat{\pi}^{\ast\ast}}\big) \, \\ \nonumber
&= \, \Phi\big((m_o - a_{\hat{\pi}^{\ast}})/s_{\hat{\pi}^{\ast}}\big) \, \approx \, \hat{F}_{\hat{\pi}^{\ast}}(m_o)~,
\end{align}
and we can estimate $\hat{m}_u$ by,
\begin{equation}
\hat{m}_u \, = \, (m_o - a_{\hat{\pi}^{\ast}}) \, \frac{s_{\hat{\pi}^{\ast\ast}}}{s_{\hat{\pi}^{\ast}}} + a_{\hat{\pi}^{\ast\ast}}~.
\label{eq:correctedmetric}
\end{equation}

\section{A statistical test to detect confounding}

In the presence of confounding, the restricted permutation null distribution will be shifted away from the baseline random guess value, and this shift can be used to informally infer the presence of confounding. Here, we present a hypothesis test to formally test the hypotheses,
\begin{align*}
H_0^c &: \mbox{the machine learning algorithm has not learned the confounding signal} \\
H_1^c &: \mbox{the machine learning algorithm has learned the confounding signal.}
\end{align*}
We adopt the sample mean of the restricted permutation null,
\begin{equation}
\bar{M}^\ast = \frac{1}{b} \sum_{i = 1}^{b} M_i^\ast~,
\end{equation}
as a test statistic, since it represents a natural measure of confounding. Note that under the null hypothesis that an algorithm has not learned the confounding signal, the restricted permutation null will have the same distribution as the standard permutation null. Hence, for large enough test sets we have that $M^\ast \approx N(a_{\hat{\pi}^{\ast\ast}} \, , \, s^2_{\hat{\pi}^{\ast\ast}})$, and our test statistic is asymptotically distributed as,
\begin{equation}
\bar{M}^\ast \, \approx \, N\left(a_{\hat{\pi}^{\ast\ast}} \, , \, \frac{s^2_{\hat{\pi}^{\ast\ast}}}{b}\right)~.
\end{equation}
Note that the variance of this null distribution depends on the number of permutations ($b$) used to generate the restricted permutation null, and gets smaller as we increase $b$. As a consequence, we can easily obtain a statistically significant result by increasing the number of permutations. In order to avoid this artifact, we replace $b$ by the number of test set samples in the computation of the p-value,
\begin{equation}
1 - \Phi\left( \frac{a_{\hat{\pi}^{\ast}} - a_{\hat{\pi}^{\ast\ast}}}{s_{\hat{\pi}^{\ast\ast}}/\sqrt{n}} \right)~.
\label{eq:confpval}
\end{equation}
By doing so, we guarantee that we will only be able to detect small confounding effects when we are truly well powered to do so. In Section 10, we report the results of a simulation study evaluating the empirical performance of the confounding test (Figure \ref{fig:simstudy}b). We observed good power to detect confounding under $H_1^c$, and well controlled type I error rates under $H_0^c$.

\section{Analytical results for the AUC metric}

It has been shown\cite{bamber1975} that, when there are no ties in the predicted class probabilities used for the computation of the $AUC$, the test statistic of the Wilcoxon rank sum test (also known as the Mann-Whitney U test), $U$, is related to the $AUC$ statistic by, $U = n_n\,n_p (1 - AUC)$, where $n_n$ and $n_p$ represent the number of negative and positive labels in the test set (see Section 2 of reference\cite{mason2002} for details). For large test sets, and under the null hypothesis that the machine learning algorithm has not learned the response and the confounding signal, this distribution can be approximated\cite{mason2002} by
\begin{equation}
U \, \approx \, N\left( \frac{n_n\,n_p}{2} \; , \; \frac{n_n\,n_p (n_n + n_p + 1)}{12} \right)~.
\label{eq:uapprox}
\end{equation}
Now, from the relation $AUC = 1 - U/(n_n\,n_p)$ it follows that,
\begin{equation}
AUC \, \approx \, N\left(\frac{1}{2} \, , \, \frac{n_n + n_p + 1}{12 \, n_n \, n_p}\right)~,
\end{equation}
so that $F_{\pi^{\ast\ast}}$ can be approximated by the above normal distribution. Following the definition in equation (\ref{eq:correctedmetricperm}), we set $F_{\pi^{\ast\ast}}(auc_u) = F_{\pi^\ast}(auc_o)$, so that,
\begin{equation}
\Phi\left( \frac{auc_u - 0.5}{\sigma} \right) \, \approx \, F_{\pi^\ast}(auc_o) \;\;\; \Leftrightarrow \;\;\; auc_u \, \approx \, \Phi^{-1}(F_{\pi^\ast}(auc_o)) \, \sigma \, + \, 0.5~,
\end{equation}
where $\Phi$ represents the cumulative density function of a standard normal random variable, and $\sigma = \sqrt{(n_n + n_p + 1)/(12 \, n_n \, n_p)}$. Now, observe that because the $AUC$ is a generalized U-statistic \cite{lehmann1951,serfling1980} it will also be asymptotically distributed as a normal random variable (even under the alternative). Hence, for large sample sizes,
\begin{equation}
F_{\pi^\ast} \, \approx \, N\left(a_{\hat{\pi}^\ast} \, , \, s_{\hat{\pi}^\ast}^2 \right)~, \hspace{0.5cm} a_{\hat{\pi}^\ast} = \hat{E}_{\hat{\pi}^\ast}[AUC^\ast]~, \hspace{0.5cm} s_{\hat{\pi}^\ast}^2 \, = \, \hat{Var}_{\hat{\pi}^\ast}(AUC^\ast)~,
\end{equation}
and the unconfounded $AUC$ score is then estimated as,
\begin{equation}
auc_u \, = \, (auc_o -  a_{\hat{\pi}^\ast}) \, \frac{n_n + n_p + 1}{12 \, n_n \, n_p \, s_{\hat{\pi}^\ast}} + 0.5~.
\end{equation}

Furthermore, under the null hypothesis that the classifier has not learned the confounding signal, it follows that the confounding null distribution for the test statistic $\bar{AUC}^\ast = b^{-1} \sum_{i=1}^{b} AUC_i^\ast$ is given by,
\begin{equation}
N\left(\frac{1}{2} \, , \, \frac{n_n + n_p + 1}{12 \, n_n \, n_p \, b}\right)~,
\label{eq:meanconf}
\end{equation}
where, as described before, we set $b = n$, where $n$ represents to the number of samples in the test set.

\section{Accounting for the confounder/response association structure in the target population}

In order to account for the confounder/response association structure in the target population we need to derive a baseline null distribution that preserves the structure observed in the target population, and use this distribution in place of the standard permutation null in our tools. For concreteness, we present next a synthetic data example describing the approach.

\begin{figure}[!t]
  \centering
  \centerline{\includegraphics[width=\linewidth]{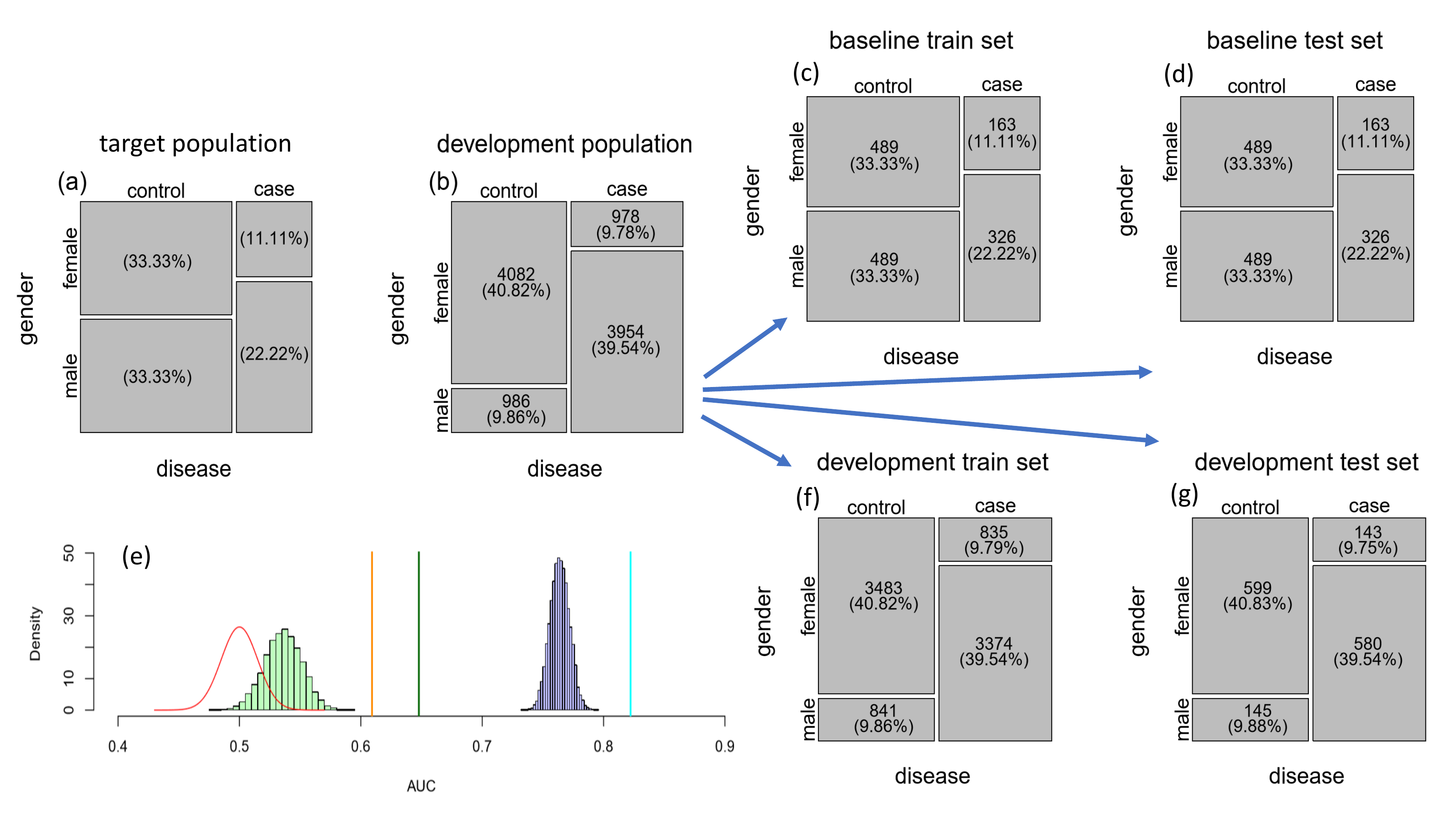}}
  \caption{An example on how to account for the confounder/response association structure in the target population. See the text for details.}
  \label{sfig:targetdevel}
\end{figure}

Suppose that it is known, a priori, that a disease affects one third of the population and is two times more common in males than in females in the target population. The mosaic plot in Figure \ref{sfig:targetdevel}a describes the joint distribution of gender and disease status in the target population. Suppose that the development set has 10,000 samples, and we are interested in building a classifier of disease status. Suppose, that due to self-selection mechanisms gender and disease status are more strongly associated in the development dataset than in the target population, as shown by the mosaic plot in Figure \ref{sfig:targetdevel}b (generated from synthetic data simulated with strong association between gender and disease status, as described in Section 10).

In order to account for the confounder/response association structure in the target population, we first need to generate a baseline null distribution that preserves this structure. To this end, we first sub-sample (from the development population) a training and test set showing the same joint distribution of gender and disease status as the target population. Figures \ref{sfig:targetdevel}c and d show the mosaic plots for these baseline training and test sets. Next, we apply restricted permutations to these subsets in order to generate the baseline null distribution (green histogram in Figure \ref{sfig:targetdevel}e), which captures the gender/response association structure of the target population. (Note how this null distribution is shifted away from 0.5, due to the association between gender and disease status.)

To quantify the amount of confounding observed in the development data (relative to the target population) we first need to generate the restricted permutation null distribution. To this end, we split the development data into training and test sets that preserve the joint distribution of the gender/disease label observed Figure \ref{sfig:targetdevel}b. (The test set, however, should have the same size as the baseline test set used to generate the baseline null (green histogram), in order to make these null distributions comparable.) Figures \ref{sfig:targetdevel}f and g show the mosaic plots for the development training and test sets, while the blue histogram in Figure \ref{sfig:targetdevel}e shows the restricted permutation null derived from these sets.

In order to compute the unconfounded predictive performance of the classifier (relative to the target population) we only need to use the baseline null distribution in place of the standard permutation null. For instance, setting $a$ and $s$ to represent the mean and standard deviation of the baseline null (green histogram in Figure \ref{sfig:targetdevel})e, and letting $a_{\hat{\pi}^{\ast}}$ and $s_{\hat{\pi}^{\ast}}$ represent, as before, the mean and standard deviation of the restricted permutation null (blue histogram in Figure \ref{sfig:targetdevel}e), we can estimate the unconfounded AUC value as,
\begin{equation}
(AUC_o - a_{\hat{\pi}^{\ast}}) \frac{s}{s_{\hat{\pi}^{\ast}}} + a~.
\label{eq:auccorrectionformula2}
\end{equation}
The green line in Figure \ref{sfig:targetdevel}e represents the above estimate (while, for the sake of comparison, the orange one shows the estimate with respect to the standard null distribution). Similarly, we can still test for the presence of confounding (which in this example is measured by the amount to association between the features and response that goes beyond the association present in the target population). To this end, we can use the $N(a , s^2/n)$ distribution as an approximate null and compute the p-value for the confounding test as,
\begin{equation}
1 - \Phi\left( \frac{a_{\hat{\pi}^{\ast}} - a}{s/\sqrt{n}} \right)~.
\label{eq:confpval2}
\end{equation}
Note that the estimator in equation (\ref{eq:auccorrectionformula2}) corresponds to formula (\ref{eq:correctedmetric}), while the p-value in equation (\ref{eq:confpval2}) corresponds to the p-value in (\ref{eq:confpval}) with $a_{\hat{\pi}^{\ast\ast}}$ and $s_{\hat{\pi}^{\ast\ast}}$ replaced by $a$ and $s$.

\section{Simulation experiments}

\begin{figure}[!h]
  \centering
  \centerline{\includegraphics[width=4.85in, bb = 0 200 1000 530]{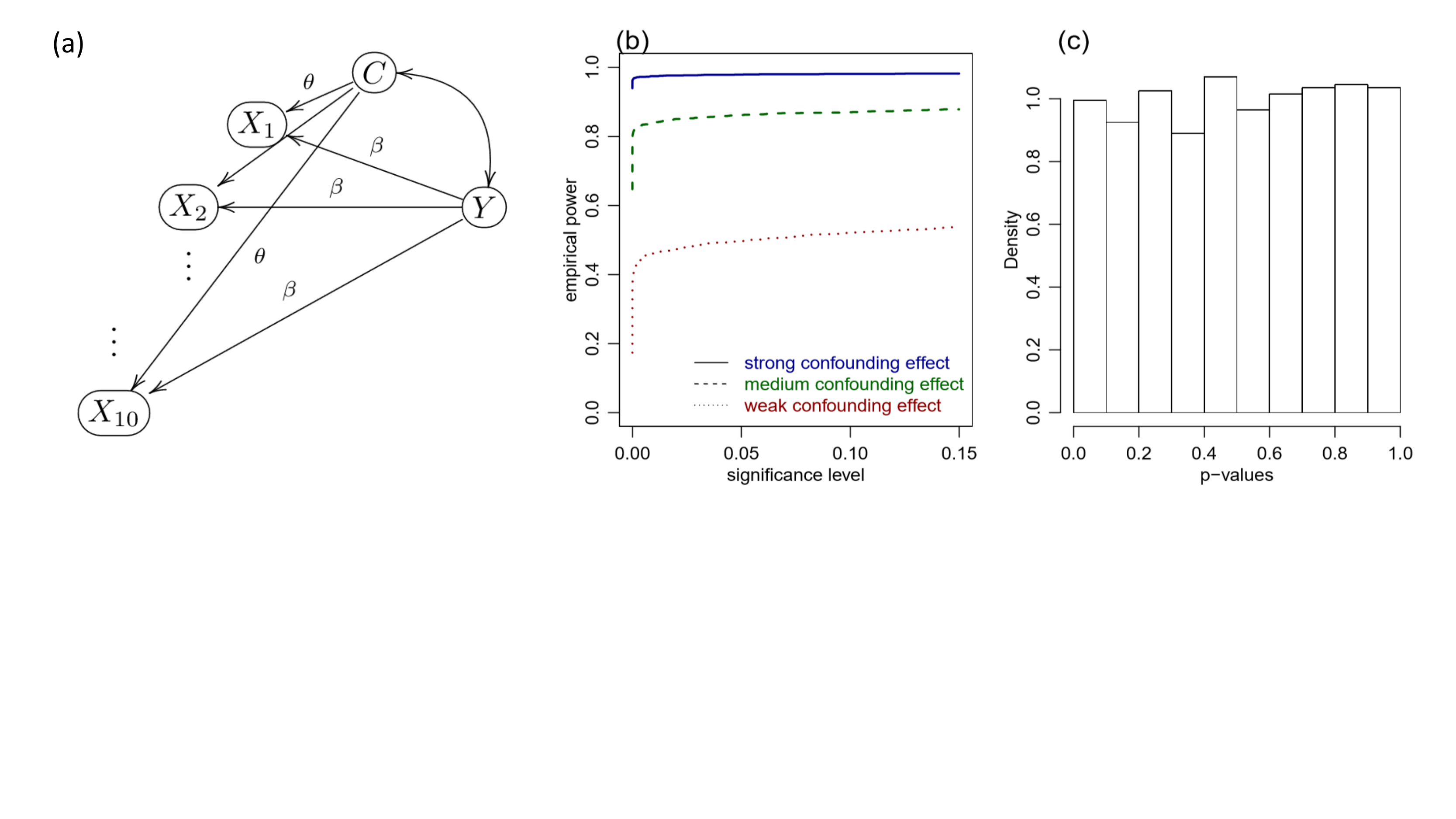}}
  \caption{Panel a describes the data generating process model. Panel b shows the empirical power curves for data simulated under the alternative hypothesis. Panel c shows the distribution of the p-values for data simulated under the null hypothesis.}
  \label{fig:simstudy}
\end{figure}

Here, we investigate the statistical power and type I error rates of the confounding statistical test ($H_0^c$ vs $H_1^c$). We simulated data according to the model in Figure \ref{fig:simstudy}a, where $C$ represents a binary confounder, $Y$ represents the disease status, and $X_1$, \ldots, $X_{10}$ represent the features. In order to generate an association between $C$ and $Y$ (i.e., $C \leftrightarrow Y$) we jointly sample these binary variables from a bivariate Bernoulli distribution\cite{dai2013}. We performed several simulation experiments based on data generated with: disease and confounding signal; disease but no confounding signal; no disease or confounding signal; and confounding but no disease signal. In each experiment we generated 1,000 data sets.

Figure \ref{fig:simstudy}b reports the empirical power curves for data simulated under $H_1^c$ (both in the presence and in the absence of disease signal) using strong, moderate, and weak amounts of confounding signal. To estimate the empirical power we recorded the proportion of times that we rejected the null hypothesis across a grid of nominal significance levels varying from 0 to 0.15. As expected, the empirical power to detect confounding increased with the strength of the confounding signal. Figure \ref{fig:simstudy}c reports the distribution of the confounding test p-values for data simulated under the null hypothesis $H_0^c$ (both in the presence and in the absence of disease signal). As expected, the distribution is close to the uniform distribution in the $[0, 1]$ interval, showing well controlled type I error rates.

\end{document}